\documentclass[preprint,authoryear,12pt]{elsarticle}

\usepackage{amsmath,amssymb}

\newcommand{\nl}{n_{\rm lock}}

\newcommand{\fb}{$f_{\rm beat}$~}

\newcommand{\fbi}{$f^{(1)}_{\rm beat}$~}
\newcommand{\fbim}{f^{(1)}_{\rm beat}}
\newcommand{\fbii}{$f^{(2)}_{\rm beat}$~}
\newcommand{\fbiim}{f^{(2)}_{\rm beat}}
\newcommand{\dfbm}{\Delta f_{\rm beat}}

\newcommand{\fr}{$f_{\rm rep}$ }
\newcommand{\frm}{f_{\rm rep}}
\newcommand{\frim}{f^{(1)}_{\rm rep}}
\newcommand{\friim}{f^{(2)}_{\rm rep}}

\newcommand{\fo}{$f_{0}$~}
\newcommand{\fom}{f_{0}}
\newcommand{\dfo}{$\Delta f_0$~}
\newcommand{\dfom}{\Delta f_0}
\newcommand{\foi}{$f^{(1)}_{0}$~}
\newcommand{\foii}{$f^{(2)}_{0}$~}
\newcommand{\foim}{f^{(1)}_{0}}
\newcommand{\foiim}{f^{(2)}_{0}}

\begin{document}

\begin{frontmatter}



\title{Optical phase-noise dynamics of \\ Titanium:sapphire optical frequency combs}


\author{Qudsia Quraishi,$^{1,4*}$ Scott Diddams,$^2$ and Leo Hollberg$^3$}

\address{
$^1$ Army Research Laboratory, Adelphi, MD 20783
\\
$^2$National Institute of Standards and Technology, Boulder, Colorado 80305\\
$^3$Department of Physics, Stanford University, Stanford, California 94305 \\
$^4$Formerly with Department of Physics, University of Colorado, Boulder, CO 80309 \\
$^*$Corresponding author: qudsia.quraishi@gmail.com
}

\begin{abstract}
Stabilized optical frequency combs (OFC) can have remarkable levels of coherence across their broad spectral bandwidth. We study the scaling of the optical noise across hundreds of nanometers of optical spectra. We measure the residual phase noise between two OFC's (having offset frequencies \foi and $\foiim$) referenced to a common cavity-stabilized narrow linewidth CW laser. Their relative offset frequency \dfo = \foii - $\foim$, which appears across their entire spectra, provides a convenient measure of the phase noise.  By comparing \dfo at different spectral regions, we demonstrate that the observed scaling of the residual phase noise is in very good agreement with the noise predicted from the standard frequency comb equation. 
\end{abstract}


\end{frontmatter}

The development of stabilized optical frequency combs (OFC) allows for straightforward schemes for comparisons of optical frequencies which are separated by 100`s of terahertz. When OFC's are referenced to stable atomic transitions, the OFC provides a bridge to evaluate relative instabilities based on different atomic frequency standards \cite{Diddams01,Gerginov06,Ma04,Newbury11}. In such an atomic-optical `clockwork', the fractional instability of the (stabilized) OFC must be sufficiently low that it does not degrade the measurement of the stability of the atomic transition. In fact, recent work has shown that the frequency instability of OFC can achieve levels of 10$^{-19}$ at averaging times of 500 seconds (when compared against another stabilized OFC \cite{Ma04}). Building on previous work with Titanium:sapphire \cite{Bartels04, Schibli08, Stenger02}, which showed phase-coherence across the optical spectrum, here, we focus on the optical phase-noise dynamics on shorter times scales (100 ns $\leq \tau \leq$ 1 s). We explore factors that limit the noise floor and demonstrate that the measured scaling of the phase-noise exhibits the scaling expected from the simple frequency comb equation.  

\begin{figure}
	\centering
\includegraphics[width=110mm,height=100mm]{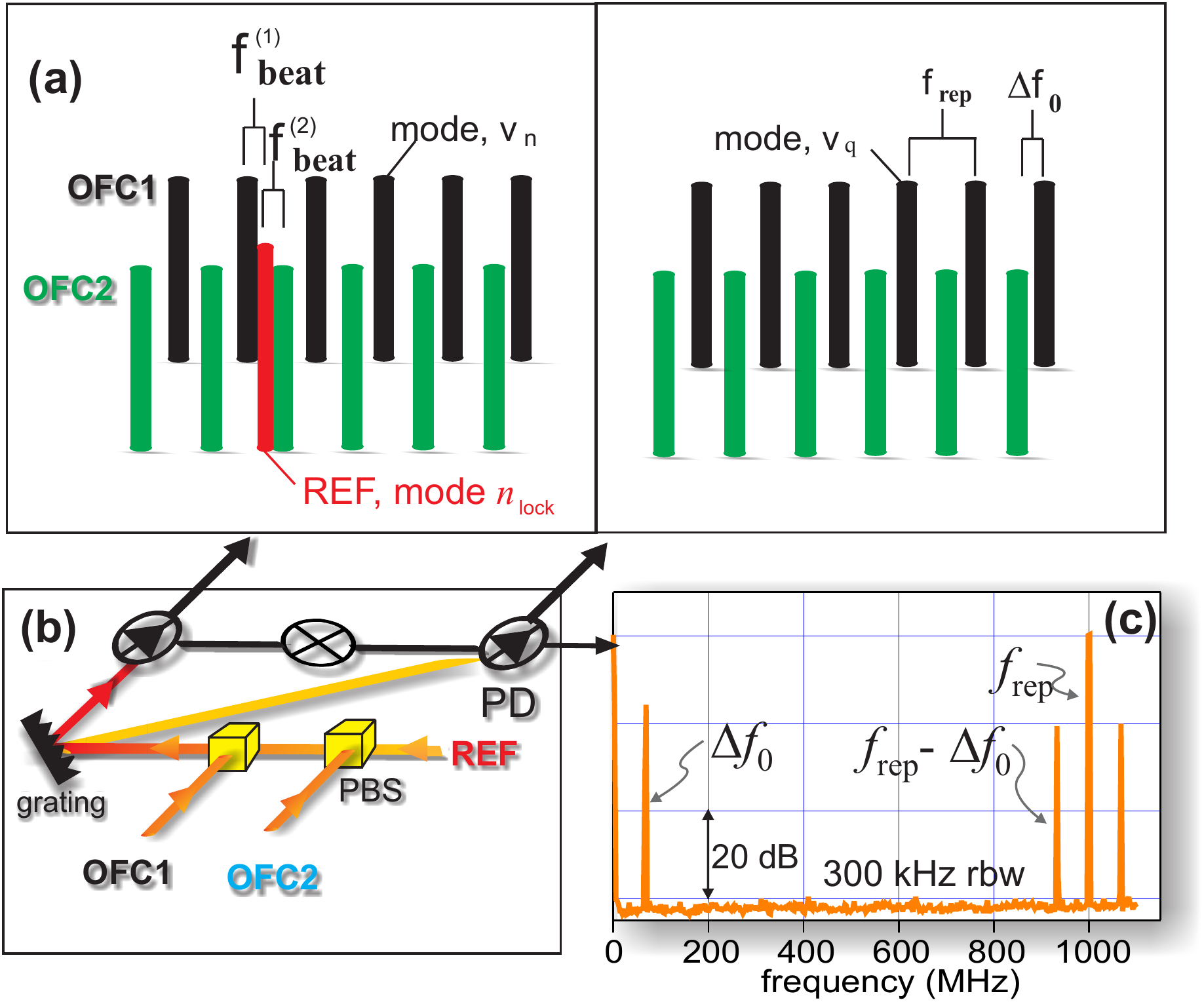}
	\caption{(a) Schematic of the optical frequency comb (OFC) output of two 1 GHz repetition rate TiS mode-locked lasers. One optical mode of each OFC is referenced, using a phase-locked loop (PLL), to a cavity-stabilized narrow linewidth CW source (REF) at 657 nm. (b) Schematic of the setup used to compare \dfo = \foii - \foi derived from two spectral regions, $\nu_n$ and $\nu_q$. Polarizing beam splitters (PBS) and a waveplate (W) are used to obtain spatial and polarization overlap. (c) Photodetected heterodyne beat of $\dfom$ from one spectral region.}
	\label{Fig1schematic}
	\!\!\!\!\!\!\!\!\!\!
\end{figure}

To determine the optical phase-noise attributable to the OFCs we lock two independent OFCs to the same cavity-stabilized CW optical frequency reference and measure the residual phase-noise between the two combs [Fig.~\ref{Fig1schematic}(a)].  Alternatively, one could use two highly-stabilized optical frequency references to measure the phase fluctuations of individual comb modes. That approach would add a source of phase noise not attributable to the combs. Such a scheme is used for comparisons of optical atomic clocks \cite{Rosenband08}, spectroscopy \cite{Ye12} and for frequency comparisons \cite{Coddington07}, in which high levels of long-term coherence have already been demonstrated. While comparing two free-running frequency combs \cite{Keller07} give some measure of phase-noise, locking the frequency combs allows one to robustly quantify the noise across a broad spectral region and compare with theoretical predictions. The excellent stability of the comb \cite{Ma04} makes it an excellent reference tool in spectral regions from microwave and terahertz to optical domains \cite{Quraishi05}.

Our apparatus consists of passive two mode-locked titanium:sapphire (TiS) ring lasers having a pulse repetition rate of 1 GHz. These self-referenced lasers have been discussed in detail elsewhere \cite{Bartels04} and employ piezoelectric actuators for cavity repetition rate stabilization.  The offset frequency of both OFC's stabilized with an 2f-to-3f technique \cite{Ramond02} which requires less than an octave of optical bandwidth, in our case, from approximately 600 nm to 1150 nm. The $n^{\rm th}$ optical frequency mode is identified as 

\begin{equation}
\!\!\!\!\!\!
\nu_n = n\frm + \fom   
\label{freqcomb_eqn}
\end{equation}

\noindent
where \fr is the repetition rate and \fo is the offset frequency \cite{Udem02} and $n$ indexes an optical mode and is an element of the integers of order $10^5$. However, when an OFC is referenced to a stable optical source with frequency $\nu_0$ (at 657nm), we can also identify the optical mode as $\nu_n = \nu_0 + f^i_{\rm beat}$, where the RF heterodyne beatnote between a mode from OFC1 (OFC2) and the CW reference is denoted by $\fbim$ ($\fbiim$) or $f^i_{\rm beat}$ (where $i=1,2$).  Solving for the repetition rate we have, $f^i_{\rm rep} = (\nu_0 + f^i_{\rm beat} - f^i_{0})/n$.

\begin{figure}
	\centering
\includegraphics[width=90mm,height=60mm]{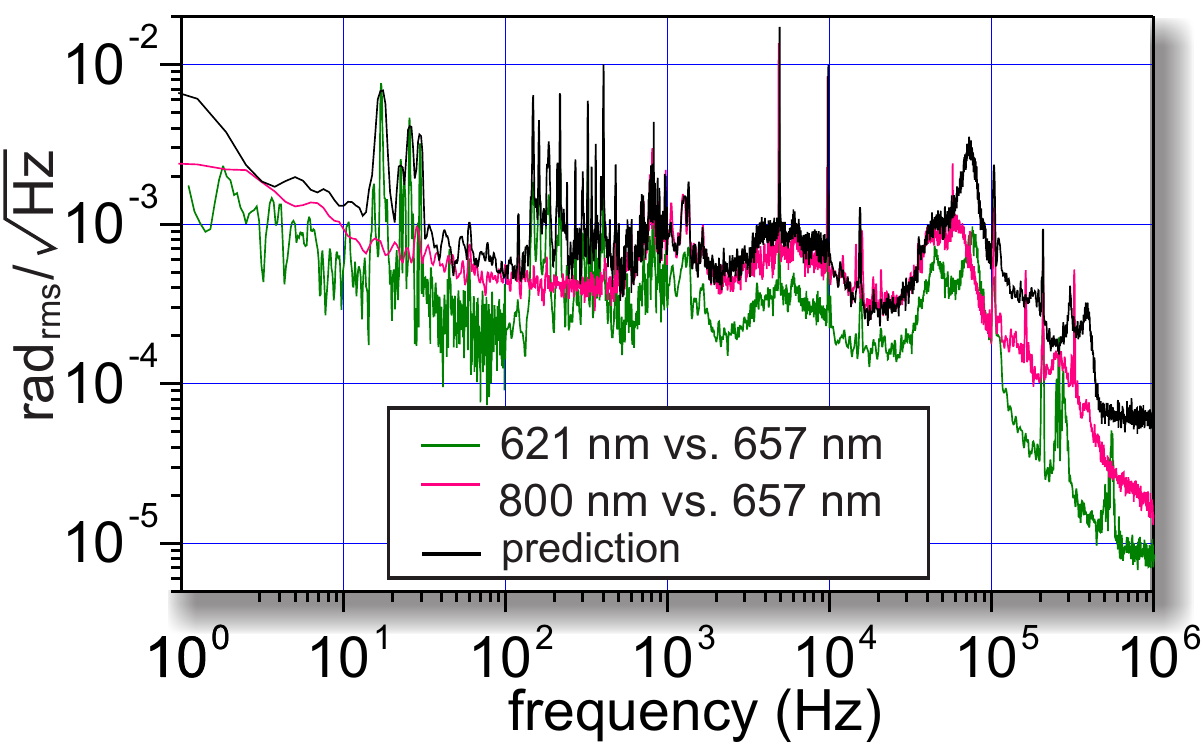}
	\caption{Measured residual phase noise (double sideband) between \dfo from two spectral regions, 800 nm and 621 nm (bandwidth $<$10 nm) as compared to the spectral region at 657 nm (bandwidth $<$3 nm), the left axis units are rad/$\sqrt{\rm Hz}$. The prediction is for the residual phase noise at 800 nm, obtained from Eqn.~\ref{psd_eqn}.}
	\label{Fig2Noise}
		\!\!\!\!\!\!\!
\end{figure} 

When we tune the repetition rates of two OFC's to be equal, we have a uniform frequency shift (\dfo $=\foiim - \foim$) between all of the modes of the two combs [Fig.~\ref{Fig1schematic}(a)]. Here, we are interested in the scaling of the phase noise away from the optical lock point at 657 nm (that is, from an imposed fixed point \cite{Telle05} of the comb). At the lock point, we measure \dfo within a small bandwidth about 657 nm and then compare it with \dfo measured at another spectral region [Fig.~\ref{Fig1schematic}(b)]. We observe excellent signal-to-noise ratios (SNR) on the \dfo signal because 1 nm of optical bandwidth corresponds to $\sim$1000 modes contributing to the measured beatnote. Figure~\ref{Fig1schematic}(c) shows the measured RF heterodyne signal from OFC1 and OFC2 which are spatially, spectrally and temporally combined. 

To calculate the residual phase noise, we begin by expressing each optical mode in terms of the laser's free parameters, $f_{0}$ and $f_{\rm beat}$, to obtain: $\nu^i_n = r_n(\nu_0 + f^i_{\rm beat}) + (1-r_n)f^i_{0}$ (where $r_n=n/\nl$ and $\nl$ indexes the mode nearest the CW optical reference) \cite{Newbury07}. The  optical phase-noise signal between OFC1 and OFC2 for the spectral region $n$ may be expressed as $\delta\nu_n = \nu^{(2)}_n-\nu^{(1)}_n$,

\!\!
\begin{equation}
\delta\nu_n = r_n(\dfbm -  \dfom) +  \dfom = r_n \Delta f_{\rm rep} + \dfom 
   \label{eqn1}
\!\!\!\!\!\!\!\!\!\!\!\!
\end{equation}

\noindent
where $\Delta f_{\rm rep} = \friim - \frim$ = 0. A similar equation may be written for the spectral region around the lock point at 657 nm, and this is denoted by $\delta\nu_q$. The relative phase noise power spectral density (PSD) $ S_{\Phi}(f) $ away from the lock point is calculated from the variance of $\delta\nu_n - \delta\nu_q$, scaled with respect to the noise bandwidth, giving

\!\!\!\!\!\!\!\!\!
\begin{eqnarray}
S_{\Phi,n,q}(f) = (r_n - r_q)^2[S_{\Phi,\fbiim}(f) + S_{\Phi,\foiim}(f) \nonumber & & \\
 + S_{\Phi,\fbim}(f) +  S_{\Phi,\foim}(f)],&&
 \label{psd_eqn}
\end{eqnarray}

\noindent
where $S_{\Phi,\fbim}(f)$ and $S_{\Phi,\foim}(f)$ denote the PSD of the electronic locks and where we have neglected all cross terms [such as $S_{\Phi,f^i_{0}}(f)\bigotimes S_{\Phi,f^j_{\rm beat}}(f)$]. Phase noise on the relative offset frequency \dfo is a measure of the combs' residual optical phase noise. The residual phase noise is the remaining phase noise between the two combs after common mode noise has been subtracted. Notably, both OFCs are locked to a common CW reference to better ensure that noise on the beat between the two OFCs (Eqn.~\ref{eqn1}) yields noise of one comb with respect to the other and not simply a measure of different CW references.  Regarding coupling between an individual OFC's two parameters of  $f^i_0$ and $f^i_{beat}$, we note that the two OFCs have independent pump lasers and phase-locked loops so ideally there should be little correlation between the two noise sources on each OFCs. Some coupling between $f^i_0$ and $f^i_{beat}$ is evident and enhanced by amplitude noise on the pump source of the Ti:S modelocked laser. Phase-locking \fo with an acousto-optic or electro-optic transducer in the pump beam path significantly reduces the coupling \cite{Quraishi_thesis}. 

\begin{figure}
	\centering
\includegraphics[width=100mm,height=60mm]{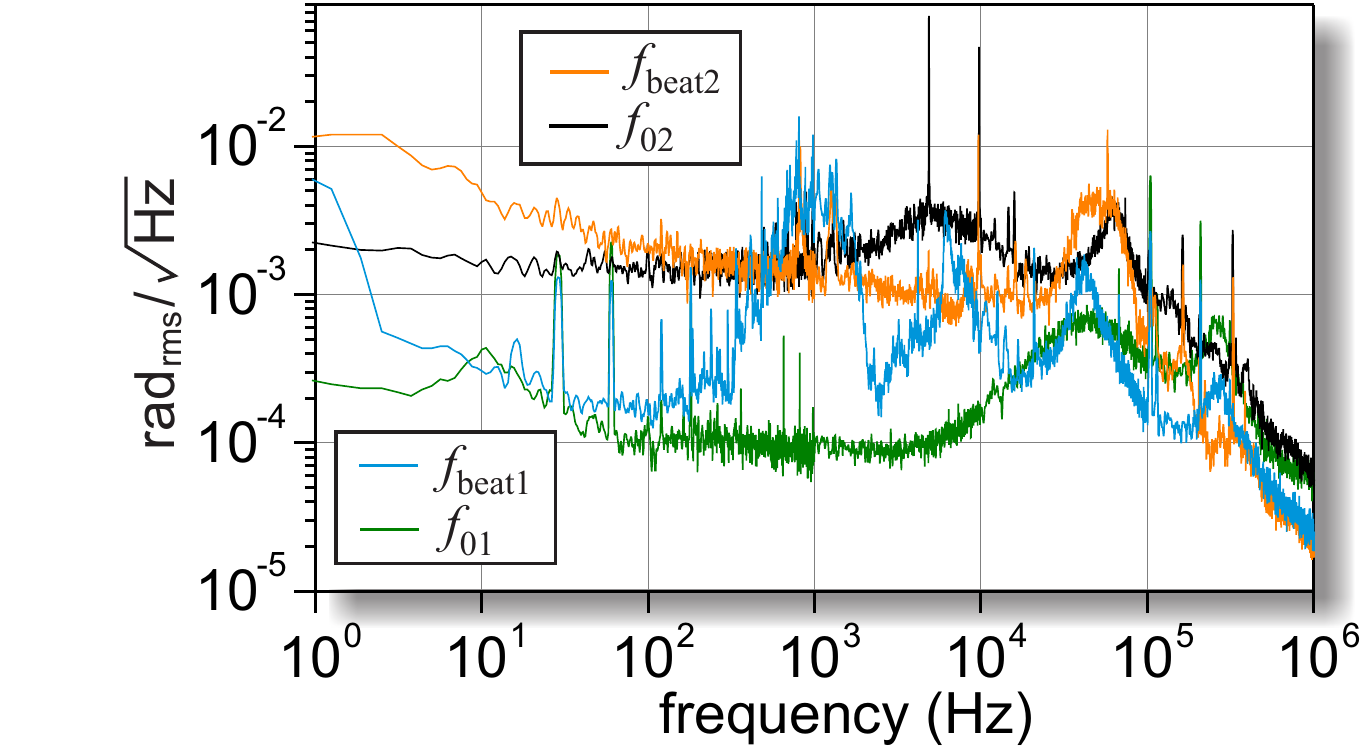}
	\caption{In-loop phase noise of the phase-locks of each optical frequency comb obtained by comparing each beatnote against a synthesizer (left axis units, rad/$\sqrt{\rm Hz}$). The noise from the offset frequencies \foi and \foii as well as the noise of the comb mode locked to the optical reference are shown (each is filtered by a 1 MHz low pass). The integrated phase on \fbi ($\foim$) corresponds to 0.11 fs (0.074 fs) of timing jitter. The integrated phase noise for the phase-locks of OFC2 for \fbii ($\foiim$) is 0.30 fs (0.15 fs).}
	\label{Fig3Locks}
		\!\!\!\!\!
\end{figure}

We measure the relative optical phase noise across the comb (Eqn.~\ref{psd_eqn}) by using the setup shown in Fig.~\ref{Fig1schematic}(b) to extract \dfo from the spectral region around 800 nm and compare that against \dfo extracted from near the lock point at 657 nm.  We observe that the optical phase noise increases with increasing frequency range away from the lock point (Fig.~\ref{Fig2Noise}).  The data in Fig.~\ref{Fig2Noise} is integrated from 1 Hz to 1 MHz, where we obtain 1.4 radians for the 800 nm data (the lock point contribution to this value is 0.9 radians). In order to obtain high resolution across all the Fourier frequencies from 0 Hz to 1 MHz, the phase noise traces shown in this work were taken by stitching together traces from smaller frequency spans where the spans initially extended less than 1 kHz and ended at a 1 MHz span.

We can also simply measure the noise of each lasers' lock to the reference separately (left hand side of Eqn.~\ref{psd_eqn}). Once we measure $f^i_0$ and $f^i_{beat}$, we can use Eqn.~\ref{psd_eqn} to predict the phase noise of comb modes ($\delta\nu_n$) away from the lock point at 657 nm. Using the results shown in Fig.~\ref{Fig3Locks} for the in-loop PSD for all the phase-locks, we plot one typical predicted phase noise in Fig.\ref{Fig2Noise}. The predicted phase noise (Fig.\ref{Fig2Noise}) is in good agreement with the measured spectral densities across the frequency range. We find comparable agreement for all of our data sets. Discrepancies arising above 55 kHz are due to the prevalence of noise outside our PLL bandwidth.

Comparing Fig.~\ref{Fig2Noise} and Fig.~\ref{Fig3Locks}, it is clear that features in the relative noise traces can be ascribed to features in each of the phase-locks. For instance, we see that \fbi ($\foiim$, $\fbiim$) dominates the noise in the vicinity of 1 kHz (5 kHz, 46 kHz).  Equation \ref{psd_eqn} gives equal relative weight to contributions from each of the phase-locks. The phase noise profiles provide useful information about the laser's phase-noise dynamics \cite{Paschotta06}. Not surprisingly, tighter phase-locks would lead to reduced optical phase noise (\dfo is a measure of the optical phase noise). The two lasers having differing noise profiles seen in Fig.~\ref{Fig2Noise} and Fig.~\ref{Fig3Locks} because they are pumped with different pump sources, have different electronic locking bandwidths and are situated in different parts of the laboratory.

Ideally, the phase noise would be reduced to the level of the shot noise, which is already the case for frequencies above $\sim$1 MHz. The majority of the noise we observe is in the frequency range below 1 MHz range. In Ti:sapphire lasers there is a strong correlation between the pump laser's amplitude noise and the optical phase noise, particularly on \fo \cite{Holman03-2}. Indeed, for the data presented in Fig.~\ref{Fig3Locks}, the phase noise is relatively high near 50 kHz due to increased amplitude noise present on the pump source of both OFC's \cite{Quraishi_thesis}. However, other noise contributions are also important including: photodetection noise (which can limit our SNR), electrical noise (ground loops, in-loop electronic noise), vibrational effects (prominent near 30 Hz), servo system gain-bandwidths (60 kHz, limited by the PZT for locking \fb\!\!) and thermal fluctuations (both within the Ti:S laser cavities and in their non-common-mode optical paths between the lasers).  Excessive noise on \fo causes optical phase noise and hence, can effect the \fb lock. 

\begin{figure}
	\centering
\includegraphics[width=90mm,height=60mm]{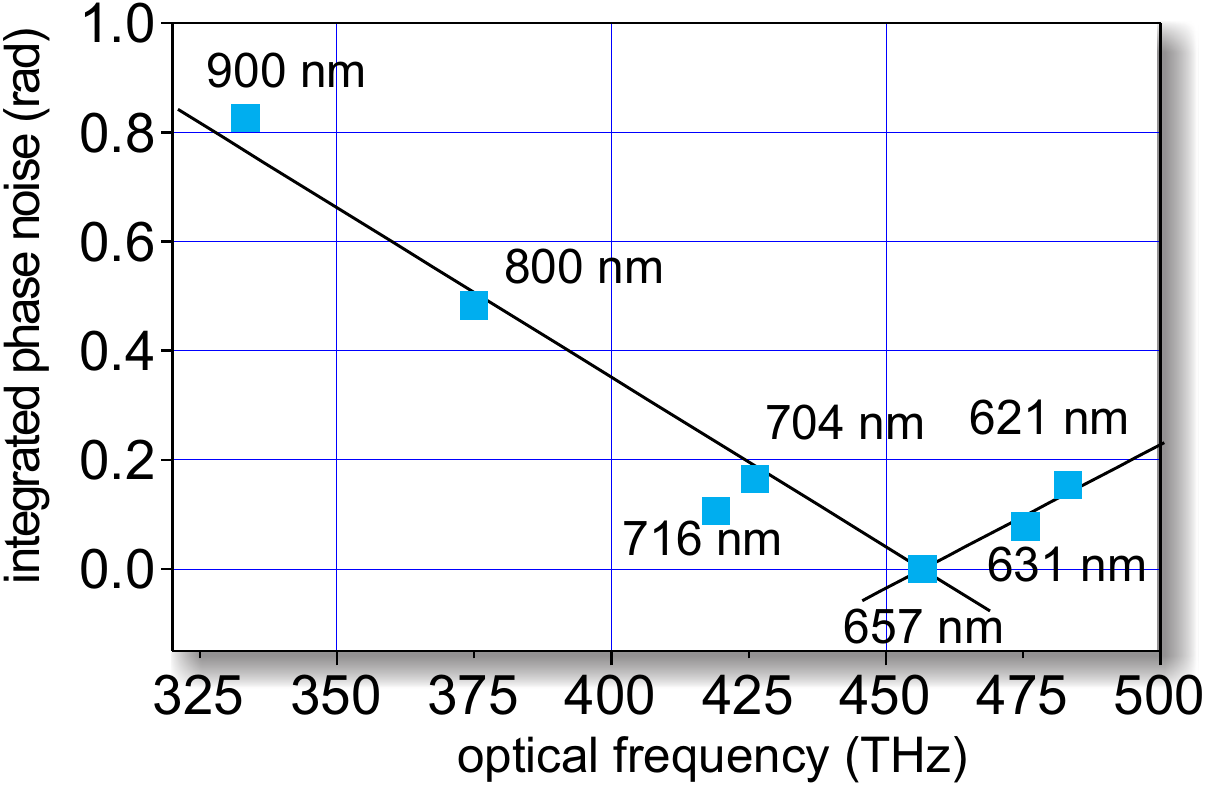}
	\caption{Scaling of the integrated (from 1 Hz to 1 MHz) phase-noise away from +243 nm (-124 THz) to -36 nm (+26 THz) from the optical lock point at 657 nm (where the lock-point phase noise of 0.9 radians is subtracted from all the data). The error bars are not visible on the scale shown. The line is a least-squares fit to the data.}
	\label{Fig4scaling}
		\!\!\!\!\!\!\!\!\!
\end{figure}

To characterize the scaling of the optical phase noise over almost 300 nm, we integrated the phase noise at discrete points in the spectrum using optical bandwidths of a few nm and RF bandwidths of 1 MHz. For each of the traces shown in Fig.~\ref{Fig2Noise} and 631 nm, 704 nm, 716 nm and 900 nm, we obtain the integrated noise values, as shown in Fig.~\ref{Fig4scaling}. Since one comb mode is locked to the reference point at 657 nm, the noise at this lock point is measured by integrating the PLL error signals of \fo and \fb. A linear trend, as predicated by Eqn.~\ref{psd_eqn}, is clearly evident [because the integrated noise $\sim [S_{\Phi,m,q}(f)]^{1/2} $ is proportional to $ (r_n - r_q)$].  A synthesizer was used to mix the lock signals to DC for the phase noise measurements and its noise level is below the measured noise levels shown here. The good agreement between the predicted and measured values demonstrate that the frequency comb equation can predict the integrated phase noise dynamics across the entire spectral range of the frequency comb.  These results were reproducible even when measurements are separated by several months.  Data for shorter wavelengths ($<$621 nm) are limited by the spectral bandwidth of our OFC (which extend down to approximately 600 nm). As seen in Fig.~\ref{Fig4scaling}, the phase noise at the lock point is 0.9 rad and increases by approximately 0.45 rad at 800 nm (meaning a culmulative phase noise value of approximately 1.35 rad). 

The 2f-to-3f technique which measures the offset frequency and locks this RF frequency, uses two vastly different spectral regions to derive a feedback signal (namely a $\sim$2 nm spectral region around both 600 nm and 900 nm). The phase noise from this lock point (\fo\!) contributes experimentally to the optical phase noise between comb modes, (see the discussion of Fig.~\ref{Fig2Noise}).  That is, when we use the frequency comb equation (Eqn.~\ref{freqcomb_eqn}) to predict the phase noise (Eqn.~\ref{psd_eqn}), as one moves away from the lock-point, we observe increased phase-noise which is well matched to the prediction, as shown in Fig.~\ref{Fig4scaling}. We see the residual phase coherence is maintained up to $\sim$30 nm away from the lock point, however we can still predict the scaling of the noise with Eqn.~\ref{psd_eqn} hundreds of nm away. This demonstrates that the frequency comb equation can also accurately predict the phase-noise. Certainly, to improve phase coherence one could use high performance phase-locks, with broadband widths and fast locking actuator feedback, such as acousto-optic modulators \cite{Koke10}. However, for applications of the frequency comb where two vastly different spectral lines are used, such as in the case of comparing optical clocks \cite{Rosenband08}, averaging the measured beat with the comb produces a highly frequency stable signal \cite{Rosenband08}, notwithstanding the lack of phase coherence across the comb. Since the scaling is predicted to be independent of the lock points (that is \fo and \fb), a future study could investigate the scaling with respect to the lock point. 

In summary, we have measured the scaling of the phase noise away from the comb's optical lock point. We observe that the majority of the measured phase noise is attributable to technical noise associated with the comb's pump source, servo system gain-bandwidths and environmental effects, which are all most significant below 1 MHz. Yet, by implementing a scheme whereby we measure the difference in phase noise between two frequency comb's locked to a common reference, we are able to measure the optical phase noise dynamics over the comb bandwidth of almost 300 nm.  The results do not show any fundamental limit to achieving shot noise performance over a broad range of the measured Fourier frequencies. Finally, we have shown that the standard frequency comb equation ($\nu_n = n\frm+\fom$) can predict not only the frequency modes of the comb but the phase noise scaling as well and we are able to measure the optical phase noise dynamics over the comb bandwidth of almost 300 nm. 

\noindent
footnote: This work is based on experiments performed in the Time and Frequency Division at the National Institute of Standards and Technology, Boulder, CO. This work is a contribution of the US government and is not subject to copyright in the US.
\!\!\!\!\!\!

\bibliography{citations}
\bibliographystyle{elsarticle-harv}



\end{document}